# Holistic Grid-Forming Control for HVDC-Connected Offshore Wind Power Plants to Provide Frequency Response

Zhenghua Xu
*DTU Wind*
*Technical University of Denmark*
Roskilde, Denmark
https://orcid.org/0000-0003-3037-4479

Dominic Groß
*Department of Electrical and Computer Engineering*
*University of Wisconsin-Madison*
Madison, USA
dominic.gross@wisc.edu

George Alin Raducu
*Power Plant Control*
*Vattenfall Vindkraft DK*
Kolding, Denmark
alingeorge.raducu@vattenfall.com

Hesam Khazraj
*Project Delivery & Engineering Power & Grid Studies (WO-PEGP)*
*Vattenfall*
Kolding, Denmark

Nicolaos A. Cutululis
*DTU Wind*
*Technical University of Denmark*
Roskilde, Denmark

*Abstract*— HVDC-connected offshore wind power plants (OWPPs) are expected to provide inertial response and frequency containment reserve (FCR) to help address the frequency control challenges caused by the growing penetration of power electronics in power systems. Initially dominated by communication-based and grid-following (GFL) control, recent efforts have shifted towards incorporating communication-free and grid-forming (GFM) control into HVDC-OWPP systems to enhance their frequency response capability. This paper proposes a holistic GFM control based on dual-port GFM control to improve the coordination across the entire AC-DC-AC dynamics. A frequency response model of a typical HVDC-OWPP system is developed for GFM control design. Then, dual-port GFM control and virtual synchronous generator control are implemented respectively on the HVDC system and OWPP of the typical system, where the asynchronism of onshore and offshore frequencies is revealed. Next, holistic GFM control is proposed to improve the synchronization and DC voltage regulation. Finally, simulation results demonstrate that the proposed control approach enables accurate frequency synchronization during FCR delivery and improved system stability during inertial response delivery.

## I. Introduction

The ongoing energy transition is driving the growing integration of converter-based renewable power generation into power systems. Consequently, system inertia is declining, and so is the available regulation reserve from conventional fossil-fuel plants as they are being replaced, making system frequency control increasingly challenging. Among the ancillary services for frequency control, inertial response is crucial to slow the rate-of-change-of-frequency (RoCoF) and buy time for the action of frequency containment reserve (FCR) which is critical to arrest frequency deviation. As an emerging large-scale power supply, HVDC-connected offshore wind power plants (OWPPs) are expected to contribute to stabilizing onshore grid frequency. Therefore, this paper investigates control approaches for providing inertial response and FCR from HVDC-connected OWPPs to the onshore grid.

The existing control approaches can be categorized into communication-based control (CBC) and communication-free control (CFC), in terms of how the offshore side senses the onshore frequency variation. For CBC [1], the onshore frequency is firstly detected at the onshore station and then transmitted through a communication line to the offshore side so that OWPPs can respond accordingly. For CFC [2], the onshore frequency deviation is incorporated into the onshore DC voltage reference, while the offshore frequency is modulated based on the DC voltage deviation measured at the offshore station, and OWPPs respond to the offshore frequency deviation instead.

Performance comparisons between CBC and CFC have been conducted in [3], [4], [5], demonstrating that CFC generally achieves more stable dynamics and better frequency deviation suppression than CBC for no communication-delay. However, these comparisons are limited to grid-following control for converters, leaving room for exploring alternative converter controls that may deliver improved performance and robustness.

Converter control serves as the subordinate control system in CBC and CFC, executing the regulation of power, voltage, and frequency, and can play a key role in dynamic performance. The choice of converter control depends on the type of converter being used. In this paper, modular multilevel converters (MMCs) are considered for both onshore and offshore HVDC stations, and the OWPPs consist of type-4 wind turbine generators (WTGs) with 2-level voltage source converters (VSCs).

The existing converter control can be generally categorized into grid-following (GFL) and grid-forming (GFM) control. On the AC terminal, AC-GFL converters fully rely on a PLL to synchronize with the AC grid while AC-GFM converters may adopt either power [6], DC voltage [7], or internal energy of MMC [8] for synchronization. On the DC terminal, DC-GFM converters regulate the terminal DC voltage whereas DC-GFL converters do not. Therefore, converter control can be further categorized into four types: AC-GFL & DC-GFM, AC-GFM & DC-GFL, dual-port GFL,

This project has received funding from the European Union's Horizon Europe Research and Innovation programme under the Marie Skłodowska-Curie grant agreement No 101073554.

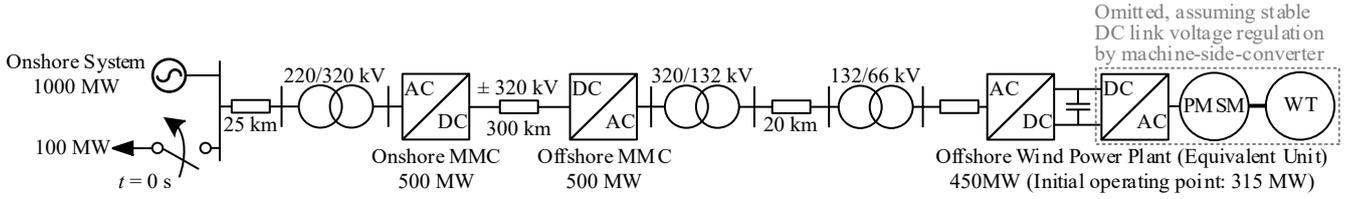
Fig. 1. A typical HVDC-OWPP system.

and dual-port GFM control [9], where AC-GFM control presents significant advantages in frequency response, achieving increased damping and lower frequency excursion, compared to AC-GFL control [10].

The converter control configuration for HVDC-OWPPs system has been rapidly evolving. Initially, both CBC [11] and CFC [12] employed AC-GFL & DC-GFM control for the onshore station, AC-GFM & DC-GFL control (specifically, *V-f* GFM control) for the offshore station, and AC-GFL control for WTGs. GFL controls break the inherent response chain in this classical configuration, causing insufficient coordination between AC and DC dynamics. Over time, researchers have been gradually introducing GFM control across the HVDC-OWPPs system to help improve frequency response. For instance, the application of GFM control to type-4 WTGs is reviewed in [13], where virtual synchronous generator (VSG) control enables WTGs to provide both FCR and inertial response inherently; [8], [14], [15] adopt various dual-port GFM controls for the onshore station, but the offshore station using *V-f* GFM control can only transfer power response passively instead of being coordinated; [16] extended dual-port GFM control to both onshore and offshore stations, whereas only the feasibility is verified without the investigation of system frequency characteristics. Besides, in [16], the onshore and offshore DC-GFM controls regulate only the DC voltage at their respective terminals, ignoring the DC line dynamics in the controller design and thus losing the coordination between DC terminals. As a result, no control approach has yet been proposed that comprehensively addresses the coordination across the entire AC-DC-AC dynamics for frequency response, from the onshore AC connection point via the HVDC system to the offshore AC grid and OWPPs' converters.

In this paper, a holistic GFM control is proposed based on CFC, dual-port GFM control for onshore and offshore MMCs, and VSG control for WTGs, aiming to improve the coordination for inertial response and FCR delivery by using GFM control on every AC and DC terminal. The rest of this paper is organized as follows: In Section II, a frequency response model of a typical HVDC-OWPP system is developed for GFM control design. In Section III, state-of-the-art GFM controls are implemented on the typical system, and the system frequency characteristics are investigated. In Section IV, the proposed holistic GFM control is elaborated. In Section V, simulations are conducted to verify the proposed control. Finally, conclusions are drawn in Section VI.

## II. SYSTEM MODELLING

A typical HVDC-OWPP system is studied in this paper, with its schematic diagram shown in Fig. 1, where the OWPP consisting of multiple WTGs is aggregated into one equivalent unit and a load event is defined as a step change for subsequent simulations.

Assuming stable control of both onshore and offshore AC voltages, the system frequency response model of the typical system is derived for GFM control design, as is presented in Fig. 2, with each component detailed as follows.

### A. Onshore Sytem

The frequency dynamics of the onshore system is modelled as an equivalent synchronous machine with swing equation and prime mover dynamics, where $H_{sys}$ and $D_{sys}$ are the system inertia and damping; $f_{sys}$ and $\theta_{sys}$ are the system frequency and phase angle; $\omega_b$ is the base value for angular frequency; $R$ is the droop coefficient for prime mover; $\Delta P_{dstb}$ is the power disturbance.

### B. Onshore and Offshore AC Grids

Thevenin equivalents are adopted for the onshore and offshore AC grids with resistance neglected, where $X_{eq, on}$ and $X_{eq, off}$ are the equivalent inductance of onshore and offshore AC grids. Without AC transmission losses, the active power transmitted through onshore and offshore AC grids, $P_{ac, on}$ and $P_{ac, off}$, can be calculated by the voltages, inductances, and phase angle differences, where $E_{sys}$, $U_{ac, on}$, $U_{ac, off}$, and $U_{OWPP}$ are respectively the terminal voltage amplitudes of the equivalent onshore system, onshore and offshore MMCs, and OWPP.

### C. Onshore and Offshore MMCs

With GFM control to be implemented, on the AC side, both onshore and offshore MMCs are modelled as a controlled AC voltage source (respectively with stable $U_{ac, on}$ and $U_{ac, off}$ and controllable output phase angles, $\theta_{MMC, on}$ and $\theta_{MMC, off}$) in series with an impedance (of MMC arms and transformer that is integrated into $X_{eq, on}$ and $X_{eq, off}$, respectively), while on the DC side, both onshore and offshore MMCs are modelled as a

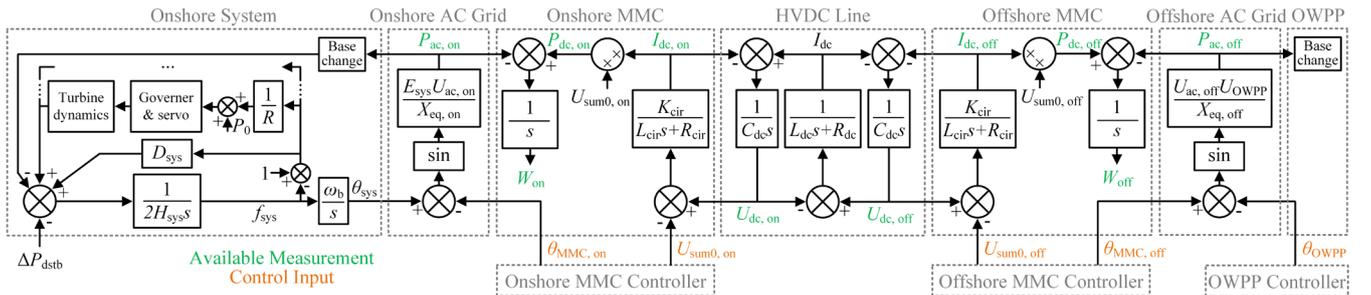
Fig. 2. System frequency response model of the typical HVDC-OWPP system with variables in per unit (p.u.)

controlled DC voltage source (driven by zero-sequence additive voltage [17], $U_{\text{sum0, on}}$ and $U_{\text{sum0, off}}$, respectively) in series with an impedance (of zero-sequence circulation circuit, $R_{\text{cir}}+j\omega L_{\text{cir}} = 2(R_{\text{arm}}+j\omega L_{\text{arm}})$, where $R_{\text{arm}}$ and $L_{\text{arm}}$ are the arm resistance and inductance of MMC; $K_{\text{cir}}$ is the gain from circulation current to DC line current, $I_{\text{dc, on/off}}$). The AC and DC sides are connected by the dynamics of internal energy [16], $W_{\text{on/off}}$, driven by AC power, $P_{\text{ac, on/off}}$, and DC power, $P_{\text{dc, on/off}} = U_{\text{sum, on/off}} I_{\text{dc, on/off}}$.

### D. Offshore Wind Power Plant

Assuming the machine-side converter regulates WTG's DC voltage [18], only the grid-side converter is modelled for WTGs. With GFM control to be implemented, following the same method for the MMCs' AC terminal, the OWPP is also modelled as an equivalent controlled AC voltage source (with stable $U_{\text{OWPP}}$ and controllable output phase angle, $\theta_{\text{OWPP}}$) in series with an impedance (of connection filter and transformer that is integrated into $X_{\text{eq, off}}$).

### E. HVDC Line

The π-section model in [19] is adopted for the HVDC line, where $R_{\text{dc}}$, $L_{\text{dc}}$, $C_{\text{dc}}$ are, respectively, the DC line resistance, inductance, and capacitance; $I_{\text{dc}}$ is the DC current flowing through the HVDC line.

## III. STATE-OF-THE-ART GFM CONTROLS

In this section, state-of-the-art GFM controls are discussed and implemented for the onshore and offshore MMCs and OWPP, and then the corresponding system frequency characteristics are investigated.

### A. Onshore and Offshore MMCs

With the internal energy storage of submodule capacitors serving as energy buffers, the MMC can control AC and DC power independently. To fully leverage this advantage, dual-port GFM control is adopted for both the onshore and offshore MMCs, aiming to coordinate all the onshore and offshore AC & DC dynamics. Specifically, the energy balancing control proposed in [9] is implemented, as illustrated in Fig. 3, where the superscript '*' denotes reference; subscript '$k$' denotes onshore or offshore side; Δ denotes deviation; $f_0$ is the nominal or initial frequency.

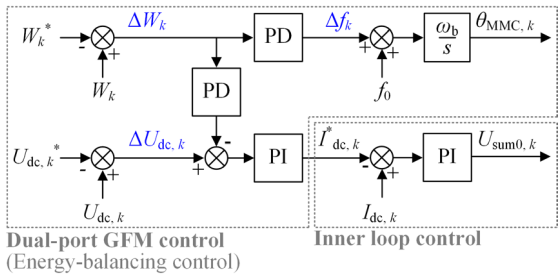

Fig. 3. Dual-port GFM controls (energy-balancing control) for onshore & offshore MMCs, $k \in \{\text{on, off}\}$.

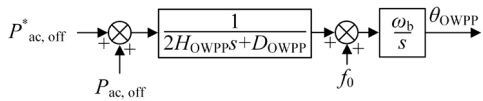

Fig. 4. Virtual synchronous generator control for offshore wind power plant.

### B. Offshore Wind Power Plant

In order to deliver an FCR and inertial response, VSG control [13] is adopted for the OWPP, as shown in Fig. 4, where $H_{\text{OWPP}}$ and $D_{\text{OWPP}}$ are the virtual inertia and damping (or droop) of OWPP.

### C. System Frequency Characteristics

According to Fig. 3, the energy-balancing control results in proportional relationships at steady-state among the deviations of frequency, internal energy, and DC terminal voltage, respectively, on onshore and offshore sides, as is shown in (1).

$$\begin{cases} \Delta f_{\text{on}} \propto \Delta W_{\text{on}} \propto \Delta U_{\text{dc, on}} \\ \Delta f_{\text{off}} \propto \Delta W_{\text{off}} \propto \Delta U_{\text{dc, off}} \end{cases} \quad (1)$$

Therefore, energy-balancing control inherently forms a CFC, where the onshore frequency variations can be transmitted through the DC voltage variations as specified in (2), where $K_1$ and $K_2$ are the proportional coefficients determined by the proportional gains of the PD controllers in Fig. 3.

$$\begin{cases} \Delta U_{\text{dc, on}} = K_1 \Delta f_{\text{on}} \\ \Delta f_{\text{off}} = K_2 \Delta U_{\text{dc, off}} \end{cases} \quad (2)$$

During the frequency event, ignoring MMC circulation loss variation, the OWPP's power response at the HVDC's offshore terminal can be expressed as (3), where subscript "0" denotes initial operating point; $U'_{\text{dc, on}} = U_{\text{dc, on, 0}} + \Delta U_{\text{dc, on}}$; $U'_{\text{dc, off}} = U_{\text{dc, off, 0}} + \Delta U_{\text{dc, off}}$.

$$\Delta P_{\text{ac, off}} = \frac{U'_{\text{dc, off}} \left( U'_{\text{dc, off}} - U'_{\text{dc, on}} \right)}{R_{\text{dc}}} - \frac{U_{\text{dc, off, 0}} \left( U_{\text{dc, off, 0}} - U_{\text{dc, on, 0}} \right)}{R_{\text{dc}}} \quad (3)$$

The system frequency characteristics regarding FCR and inertial response are further investigated below, respectively.

#### 1) FCR

According to Fig. 4, when $H_{\text{OWPP}} = 0$ (s), the power response from OWPP, $\Delta P_{\text{OWPP}}$, for FCR alone, can be expressed as (4).

$$\Delta P_{\text{OWPP}} = -\Delta P_{\text{ac, off}} = -D_{\text{OWPP}} \Delta f_{\text{off}} \quad (4)$$

By solving the set of equations (2) - (4), the offshore frequency variation can be expressed as (5), where $\beta_1 = -K_1 \Delta f_{\text{on}} + K_2 R_{\text{dc}} D_{\text{OWPP}} + 2U_{\text{dc, off, 0}} - U_{\text{dc, on, 0}}$.

$$\Delta f_{\text{off}} = \frac{K_2 \left( -\beta_1 + \sqrt{\beta_1^2 + 4K_1 \Delta f_{\text{on}} U_{\text{dc, off, 0}}} \right)}{2} \quad (5)$$

#### 2) Inertial Response

According to Fig. 4, when $D_{\text{OWPP}} = 0$ (p.u.), $\Delta P_{\text{OWPP}}$ for inertial response alone can be expressed as (4),

$$\Delta P_{\text{OWPP}} = -\Delta P_{\text{ac, off}} = -2H_{\text{OWPP}} \frac{d\Delta f_{\text{off}}}{dt} \quad (6)$$

By solving the set of equations (2), (3), and (6), the offshore frequency variation can be approximated by (7), where $\beta_2 = -K_1 \Delta f_{\text{on}} + 2U_{\text{dc, off, 0}} - U_{\text{dc, on, 0}}$.

$$\Delta f_{\text{off}} = \frac{K_2\left(-\beta_2 + \sqrt{\beta_2^2 + 4K_1\Delta f_{\text{on}}U_{\text{dc, off, 0}}}\right)}{2} \quad (7)$$

According to (5) and (7), offshore and onshore frequency deviations are not synchronized, and their nonlinear relationship depends on DC line resistance, control parameters, and initial operating point, due to the nonlinear DC power calculation and DC line losses in (3). The inconsistency between onshore and offshore frequency variations will result in incorrect delivery of FCR and inertial response services. [20] proposed that the hybrid AC/DC system using dual-port GFM control can achieve quasi-synchronous steady-state if DC line losses are negligible, implying DC compensation as a potential solution. However, while offshore $V$-$f$ GFM control can implement the DC compensation using DC current or power [2], [12] through offshore frequency modulation, dual-port GFM control cannot modulate offshore frequency directly.

## IV. HOLISTIC GFM CONTROL

The proposed holistic GFM control is shown in Fig. 5. The DC voltage references for both onshore and offshore MMCs are the same, while energy-balancing control requires different references for power dispatching. Moreover, holistic GFM control utilizes the estimation of onshore terminal DC voltage, $\hat{U}_{\text{dc\_on}}$, as offshore control feedback, considering only the influence of DC line resistance.

According to Fig. 5, holistic GFM control results in proportional relationships at steady state as shown in (8). Therefore, by properly tuning the proportional gains of the four PD controllers, holistic GFM control achieves synchronization of onshore and offshore frequencies.

$$\Delta f_{\text{on}} \propto \Delta U_{\text{on}} \approx \Delta \hat{U}_{\text{on}} \propto \Delta f_{\text{off}} \quad (8)$$

Moreover, the onshore and offshore DC regulations are unified by holistic GFM control. For energy-balancing control, as shown in Fig. 6 (a), the onshore and offshore DC voltage controllers only regulate their terminal DC voltage, respectively, while the DC current $I_{\text{dc}}$ is regarded as a disturbance and the DC line dynamics are ignored in the controller design. For holistic GFM control, as shown in Fig. 6 (b), assuming accurate estimation, the onshore and offshore DC regulations can be unified to regulate the onshore DC terminal voltage. Thereby, the DC line dynamics must be considered in the control system, and the scope of DC control design covers the entire DC dynamics.

By synchronizing onshore and offshore frequencies, unifying the regulation of DC dynamics, and exerting dual-port GFM capability, holistic GFM control can facilitate

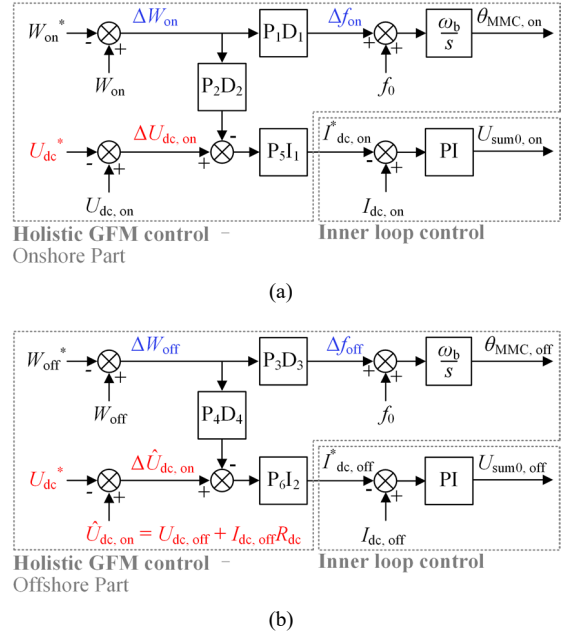

Fig. 5. Holistic GFM control.
(a) Onshore MMC; (b) Offshore MMC.

comprehensive coordination across the entire AC-DC-AC dynamics of a HVDC-OWPP system.

## V. SIMULATION RESULTS

The FCR and inertial response of the proposed holistic GFM control is verified via simulations, and its performance is compared with energy-balancing control. The models and parameters for the simulations are listed in TABLE I, where the proportional gains of holistic GFM control are tuned to achieve the synchronization of onshore and offshore frequencies ($P_2/P_1 = P_4/P_3$) and perform the same total amount of virtual inertia from onshore and offshore MMCs as energy-balancing control.

The simulation results of FCR delivery with $H_{\text{OWPP}} = 0$ (s) and $D_{\text{OWPP}} = 20$ (p.u.) are shown in Fig. 7. For energy-balancing control, as shown in Fig. 7. (a), the asynchronism results in a maximum of 8.11% discrepancy between onshore and offshore frequencies, with the offshore frequency deviation being reduced. This results in a OWPP response lower than the requirement defined by the onshore frequency variation $\Delta f_{\text{on}}$ and OWPP's droop $D_{\text{OWPP}}$ as shown in Fig. 7. (b), and the maximum discrepancy is 8.96%. For holistic GFM control, as shown in Fig. 7. (a), onshore and offshore frequencies are synchronized with little discrepancy during transients, meaning the PI controller in DC voltage control system settles (8) rapidly, and accordingly the FCR response provided by OWPP aligns well with the requirement defined by $\Delta f_{on}$ and $D_{OWPP}$, as is shown in Fig. 7. (b).

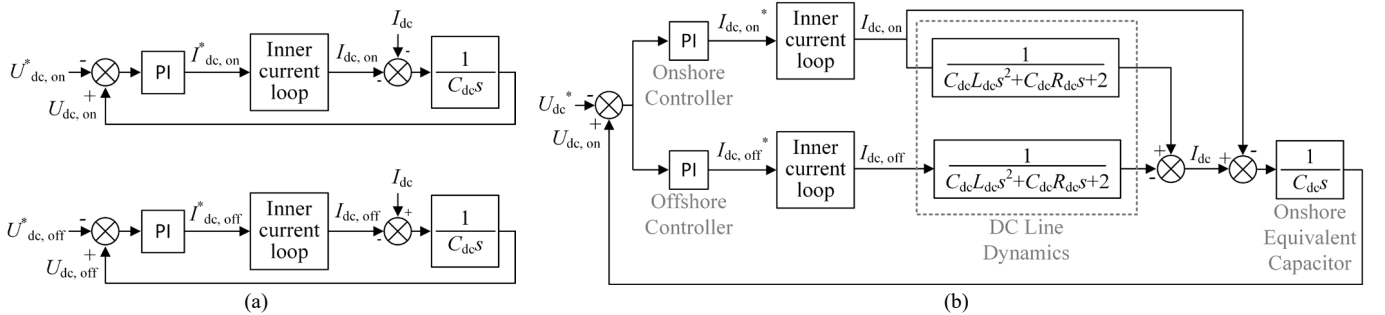

Fig. 6. DC control system. (a) Energy-balancing control; (b) Holistic GFM control.

TABLE I. MODELS AND PARAMETERS FOR THE SIMULATIONS

| Models and/or parameters | | | Sources / Values |
|---|---|---|---|
| OWPP | | | WECC type-4 wind turbine in [21] |
| MMC | | | Dual-port GFM MMC in [9] |
| Energy-balacing control | | | |
| AC & DC lines | | | |
| Synchronous generator unit ($H_{sys} = 4$ (s), $D_{sys} = 0$ (p.u.), $R = 0.05$ (p.u.) ) | | | Two-area test system in [22] |
| Holistic GFM Control | Onshore | $P_1$ (p.u.) | 1 |
| | | $D_1$ (p.u.) | 0.025 |
| | | $P_2$ (p.u.) | 1 |
| | | $D_2$ (p.u.) | 0.025 |
| | | $P_5$ (p.u.) | 11.914 |
| | | $I_1$ (p.u.) | 2382.9 |
| | Offshore | $P_3$ (p.u.) | 0.33 |
| | | $D_3$ (p.u.) | 0 |
| | | $P_4$ (p.u.) | 0.33 |
| | | $D_4$ (p.u.) | 0.025 |
| | | $P_6$ (p.u.) | 11.914 |
| | | $I_2$ (p.u.) | 2382.9 |

The simulation results of inertial response delivery with $H_{OWPP} = 4$ (s) and $D_{OWPP} = 0$ (p.u.) are shown in Fig. 8.

Using energy-balancing control, onshore and offshore frequencies are synchronized, and the deviation is not as significant as that during FCR delivery (see Fig. 8. (a) with Fig. 7. (a)). For inertial response alone, the term relating to the droop $D_{OWPP}$ is removed from the relationship between $\Delta f_{on}$ and $\Delta f_{off}$, as is shown by comparing $\beta_1$ and $\beta_2$ in (5) and (7). It is worth noting that by compensating DC losses (equivalently, $R_{dc} = 0$), the term relating to the droop $D_{OWPP}$ is also removed, which is the condition for quasi-synchronous steady-state, as proposed by [20]. Thus the close frequency synchronization observed here also verifies proposition 3 in [20]. Under holistic GFM control, the onshore and offshore frequencies are also synchronized closely during the entire transient, as shown in Fig. 8. (a), thanks again to the fast regulation of the PI controller in DC voltage control system as well as the reduced nonlinearity between $\Delta f_{on}$ and $\Delta f_{off}$ since the $D_{OWPP}$ related term is removed.

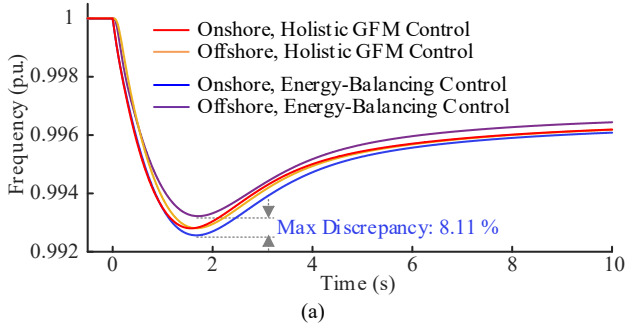

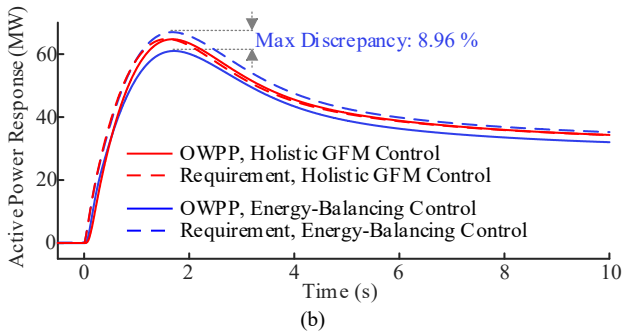

Fig. 8. Simulation results for FCR delivery with $H_{OWPP} = 0$ (s) and $D_{OWPP} = 20$ (p.u.).
(a) System frequencies; (b) Active power response from OWPP

The full synchronization between onshore and offshore frequencies enables both energy-balancing control and holistic GFM control to deliver a coherent inertial response to onshore frequency variation. This response adheres to the requirement defined by the detected onshore frequency derivative $df_{on}/dt$ and virtual inertia time constant $H_{OWPP}$, as is shown in Fig. 8. (b). Notably, the inertial responses initially oscillate around the expected response and then rapidly converge. The inertial response deviation from the requirement is highlighted in Fig. 8. (c), showing that the deviations originate from oscillations, with the oscillations under holistic GFM control being slightly lowered. Achieving an ideal inertial response from OWPPs will require further damping of the oscillations (e.g., fast MMC and DC line

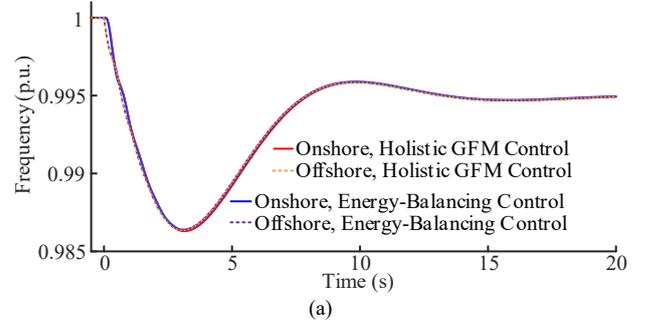

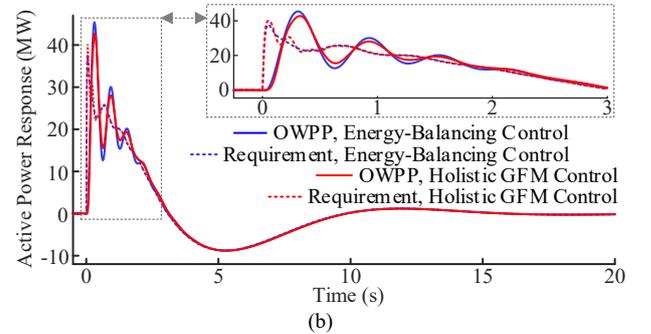

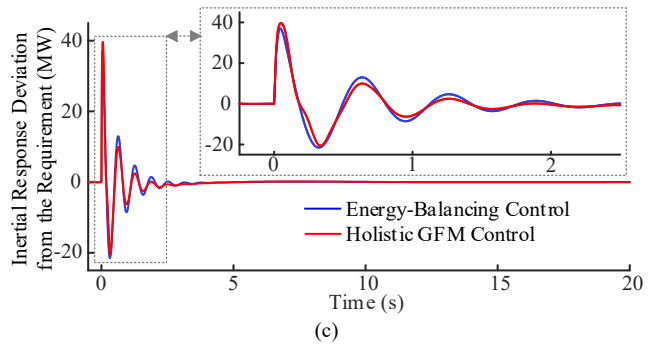

Fig. 7. Simulation results for inertial response delivery with $H_{OWPP} = 4$ (s) and $D_{OWPP} = 0$ (p.u.).
(a) System frequencies; (b) Active power response from OWPP; (c) Inertial response deviation from the standard.

dynamics), with holistic GFM control showing a relative advantage in this regard.

VI. CONCLUSION

To enhance the frequency response capability of HVDC-OWPP systems, this paper implements GFM control on every terminal of a HVDC-OWPP system (i.e., OWPP's AC terminal, HVDC system's onshore and offshore AC terminals as well as its onshore and offshore DC terminals), and analyzes challenges in synchronizing onshore and offshore frequencies.

The proposed holistic GFM control achieves steady-state frequency synchronization between onshore and offshore terminals. Furthermore, the holistic GFM control incorporates onshore and offshore DC voltage regulation to form a distributed but functionally unified control system, which ensures coherent regulation of the overall DC dynamics. In contrast, the original design regulates each DC terminal independently. By combining this unified DC voltage regulation with dual-port GFM capability and onshore & offshore frequency synchronization, holistic GFM control facilitates coordinating the overall dynamics of HVDC transmission and the connected onshore & offshore AC grids at all onshore and offshore AC & DC terminals.

Simulation results verified that the holistic GFM control closely synchronizes the onshore and offshore frequencies in steady-state and during transients, delivering accurate FCR and inertial response services. Besides, the holistic GFM control exhibits a comparative advantage in enhancing oscillation damping for inertial response.

Future work on the holistic GFM control includes: 1) Stability analysis and control improvement for oscillation damping. 2) Extending the control to 2/3-level VSC. 3) Evaluating the sensitivity of DC voltage estimation to DC line parameters and exploring influences when the estimation point is selected randomly along the HVDC line. 4) Incorporating the machine-side dynamics of WTGs into the control design. 5) Extending the control to multi-terminal HVDC grid interconnecting multiple AC grids.